\documentclass[aps,prl,showpacs, twocolumn,amsfonts,amssymb,amsmath]{revtex4}
\usepackage{graphicx} 
\usepackage{array}
\usepackage{SIunits}
\usepackage{color}
\newcommand{\YAG}[0]{Y$_2$Al$_5$O$_{12}\,$}

\newcommand{\Ce}[0]{Ce$^{3+}\:$}
\newcommand{\Fig}[0]{Fig.$\:$}
\begin{document}

\title{Measurement of line widths and permanent electric dipole moment change of the Ce 4f-5d transition in Y$_2$SiO$_5\,$ for a qubit readout scheme in rare-earth ion based quantum computing}

\author{Ying Yan, Jenny Karlsson, Lars Rippe, Andreas Walther, Diana Serrano,\\David Lindgren, Mats-erik Pistol, and Stefan Kr\"{o}ll}
\affiliation{Department of Physics, Lund University, P.O.~Box 118, SE-22100 Lund, Sweden}

\author{Philippe Goldner}
\affiliation{Chimie ParisTech, Laboratoire de Chimie de la Mati$\grave{e}$re Condens$\acute{e}$e de Paris,
CNRS-UMR 7574, UPMC Univ Paris 06, 11 rue Pierre et Marie Curie 75005 Paris, France}

\author{Lihe Zheng and Jun Xu}
\affiliation{Key Laboratory of Transparent and Opto-Functional Inorganic Materials, Shanghai Institute of Ceramics, Chinese Academy of Sciences, Shanghai 201800, China}

\date{\today }

\begin{abstract}
In this work the inhomogeneous (zero-phonon line) and homogeneous line widths, and one projection of the permanent electric dipole moment change for the Ce 4f-5d transition in Y$_2$SiO$_5\,$ were measured in order to investigate the possibility for using Ce as a sensor to detect the hyperfine state of a spatially close-lying Pr or Eu ion. The experiments were carried out on Ce doped or Ce-Pr co-doped single Y$_2$SiO$_5\,$ crystals. The homogeneous line width was measured to be about $\sim$ 3 MHz, which is essentially limited by the excited state lifetime. Based on the line width measurements, the oscillator strength, absorption cross section and saturation intensity were calculated to be about 9$\times10^{-7}$, 5$\times10^{-19}\:$m$^{2}$ and 1$\times10^{7}\:$W/m$^{2}$, respectively. One projection of the difference in permanent dipole moment, $\bigtriangleup\mu_{Ce}$, between the ground and excited states of the Ce ion was measured as \textbf{$6.3 \times10^{-30}\:$C\cdot m}, which is about 26 times as large as that of Pr ions. The measurements done on Ce ions indicate that the Ce ion is a promising candidate to be used as a probe to read out a single qubit ion state for the quantum computing using rare-earth ions.

\end{abstract}

\maketitle

\begin{center}
\textbf{I. INTRODUCTION}\\
\end{center}

The quantum computing research field has attracted extensive interest for its potential to give a tremendous boost in computational ability for certain types of problems. Many physical systems have been investigated as test beds for quantum computing \cite{Ladd}: trapped ions \cite{Blatt}, nuclei in molecules \cite{Jones}, Josephson junctions in superconductors \cite{Plantenberg}, nitrogen vacancy centers in diamond \cite{Nizovtsev}, rare-earth ions in inorganic crystals \cite{Ohlsson}, etc. Regardless of the  physical system, one of the necessary criteria for a quantum computing scheme is that it should be scalable. In the rare-earth ion based quantum computing (REIQC) approach, qubit-qubit interaction and arbitrary qubit rotations characterised by quantum state tomography have been carried out \cite{Longdell}\cite{Rippe}. In these experiments each qubit, where the qubit states ($\vert$0$>$ and $\vert$1$>$ ) are two ground state hyperfine levels of the ion, was represented by an ensemble of ions \cite{Rippe}. In the rare-earth ion doped crystals the transition lines are inhomogeneously broadened as a result of the random substitution of the rare-earth ions (qubit ions) in the solid matrix, leading to slight crystal field variations for the sites of the individual ions. The ratio between the inhomogeneous and homogeneous broadening in these systems can be larger than $10^{6}$. Thus a very large number of sub-ensembles of ions, where each sub-ensemble is selectively addressed in frequency space, can be singled out within the inhomogeneous line. Each sub-ensemble can then act as a selectively addressed qubit. For two ions in different qubits, the permanent electric dipole-dipole interaction is proportional to $1/r^{3}$, where $r$ is the the distance between the two ions. This spatially dependent coupling determines the probability, $p$, of an ion in one qubit being sufficiently close to an ion in another qubit, where it is often $p\ll1$ for reasonable dopant concentrations. This means that the number of active ions in one ensemble qubit that interact strongly with one ion in each of the rest of the $n-1$ qubits scales as $p^{n-1}$. In order to improve this poor scalability, Wesenberg \textit{et al.} proposed several schemes \cite{Wesenberg}. One approach is to discard the ensemble qubits and instead let each qubit be represented by a single ion (single instance quantum computing). With sufficient dopant concentration (about $0.2\%$ for Pr:Y$_2$SiO$_5\,$) there will always be sequences of connected ions close enough for carrying out gate operations \cite{Wesenberg}\cite{Samuel}.\\

   However in this single instance quantum computing approach a technique to read out the quantum state of a single-ion qubit needs to be developed. The straight-forward typical fluorescence measurement for detecting a single molecule does not work because: firstly the transitions of the qubit ion (In reference \cite{Rippe} this has been a Pr ion) which can discriminate the qubit states have excited state lifetimes of more than \SIunits{10} {\micro\second}, which provides too low emission rate for a high signal to noise detection; secondly and more importantly, the qubit ion candidates have more than one ground state hyperfine level to which they can decay, thus there is no transition that can be cycled until the number of emitted photons is sufficiently large to provide state selective information. A single ion readout idea was proposed in \cite{Wesenberg} to accomplish single-qubit ion detection: an additional ion (hereafter called as readout ion) can be co-doped into the crystal, with such low concentration that there is only one readout ion fluorescing within the laser focal volume. This ion serves as a sensor for reading out the state of a nearby qubit through the interaction between the qubit and readout ion, illustrated in \Fig1. As a result the fluorescence signal from the single readout ion, either ON or OFF, depends on whether the single qubit ion is in state $\vert$1$>$ or in $\vert$0$>$ \cite{Walther}, respectively. \\
\begin{figure}[h]
		\includegraphics[width=8.0cm,height=6cm]{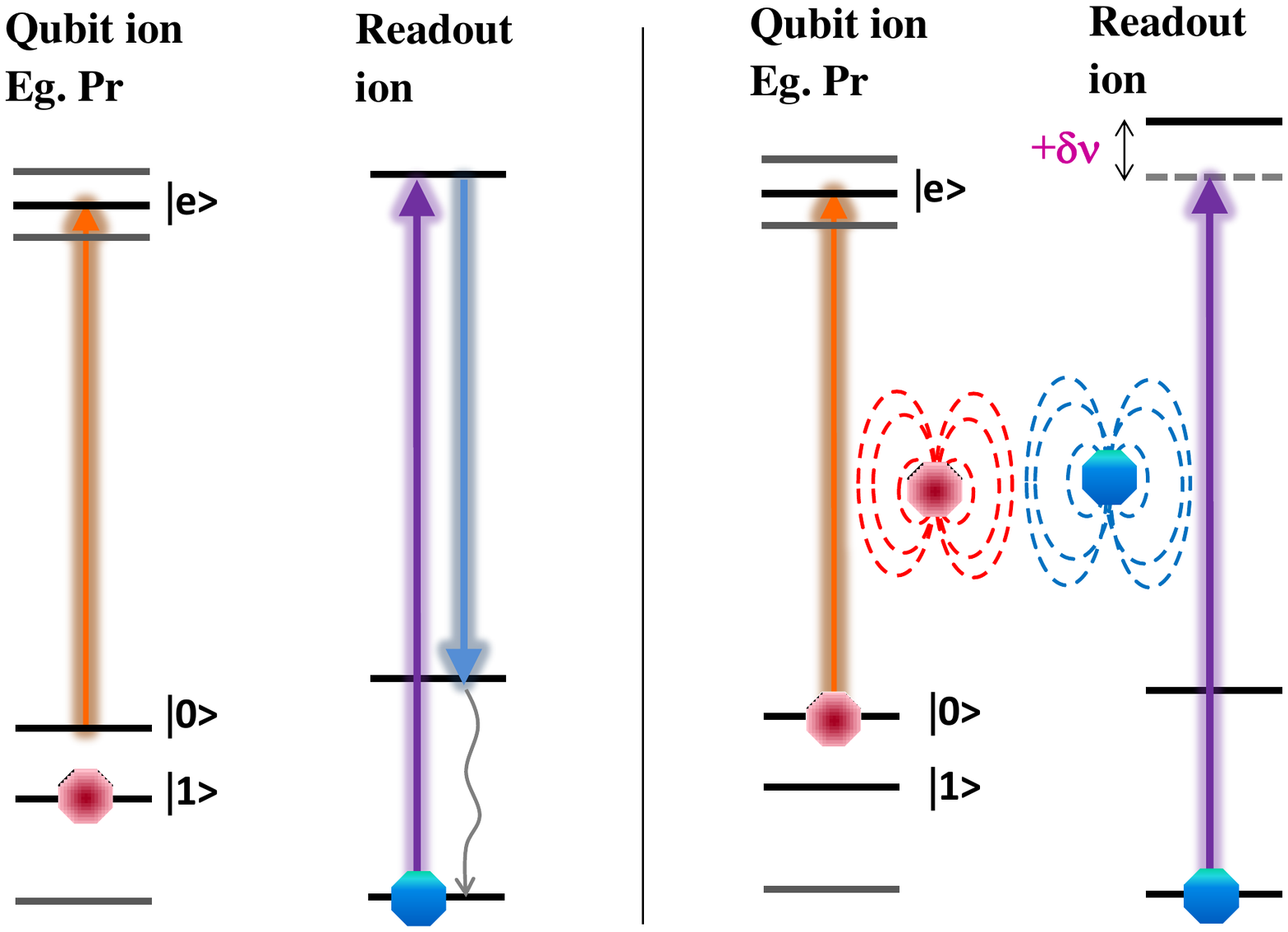}
	\caption{(color online) Permanent dipole-dipole interaction between a qubit ion and a readout ion that sits spatially close to this qubit ion. Two lasers interact with the ions. The qubit laser can send out pulses with pulse area of $\pi$ on the $\vert 0\rangle$ $\rightarrow\vert e\rangle$ transition of the qubit ion (Pr) and the readout laser continuously excites the readout ion. If the Pr ion is in the $\vert 1\rangle$ state (figure on left panel), the qubit laser is not on resonance with the Pr ion transition and the Pr ion is not transferred to its excited state. The readout laser is then resonant with the readout ion transition and the readout ion continuously sends out fluorescence photons. On the other hand, if the Pr ion is initially in the $\vert 0\rangle$ state (figure on right panel), it will be excited by the qubit laser pulse, since the permanent dipole moment of Pr ion in state $\vert e\rangle$ is different from that in state $\vert 0\rangle$, the change of this local electric field induces a frequency shift of the nearby readout ion transition line. If the shift is larger than the homogeneous line width, the readout ion will be out of resonance with the readout laser and the fluorescence is turned off.}
	\label{fig.1}
\end{figure}

The readout scheme above requires that the readout ion has the following characteristics. (i) Short excited state lifetime compared to the qubit ion lifetime in order to get a large number of fluorescence photons while the qubit is in the excited state. (ii) Narrow homogeneous absorption line width such that the permanent dipole-dipole interaction with a nearby qubit shifts the readout ion resonance frequency by several homogeneous line widths. (iii) Large dipole moment change between ground and excited state (again such that the shift due to the permanent dipole-dipole interaction is sufficiently large). (iv) No fluorescence quenching mechanisms, \textit{e.g.} a long-lived trapping state or energy transfer from the readout ion to the qubit ion. In this work the Ce ion (doped in an Y$_2$SiO$_5\,$ crystal) is considered as a readout ion \cite{Olivier}. The Ce ion in Y$_2$SiO$_5\,$ has a short excited state lifetime of abut 50 ns \cite{Aitasalo}\cite{Julio}, to be compared with an excited state lifetime of possible qubit ions as \textit{e.g.} Pr and Eu which are about 0.2 ms and 2 ms, respectively. The 4f-5d zero-phonon absorption line of \Ce doped in Y$_2$SiO$_5\,$ lies around 370.83 nm, which is well separated from the qubit transition frequencies (\textit{e.g.} 606 nm for Pr ions in site 1). However, other than the excited state lifetime, the spectroscopic parameters relevant for the read out scheme are not known. In this article, the second and third spectroscopic requirements, (ii) and (iii) above, for a readout ion were measured for the Ce ion in an Y$_2$SiO$_5\,$ crystal. The inhomogeneous zero-phonon line (ZPL) measurement is described in Section II A, the homogeneous line width measurement is discussed in Section II B, and the measurement of the Ce ground and excited state permanent dipole moment difference (based on the Ce-Pr interaction) is described in Section II C. The work is concluded in Section III.\\

\begin{figure}[h]
		\includegraphics[width=8.5cm,height=5.5cm]{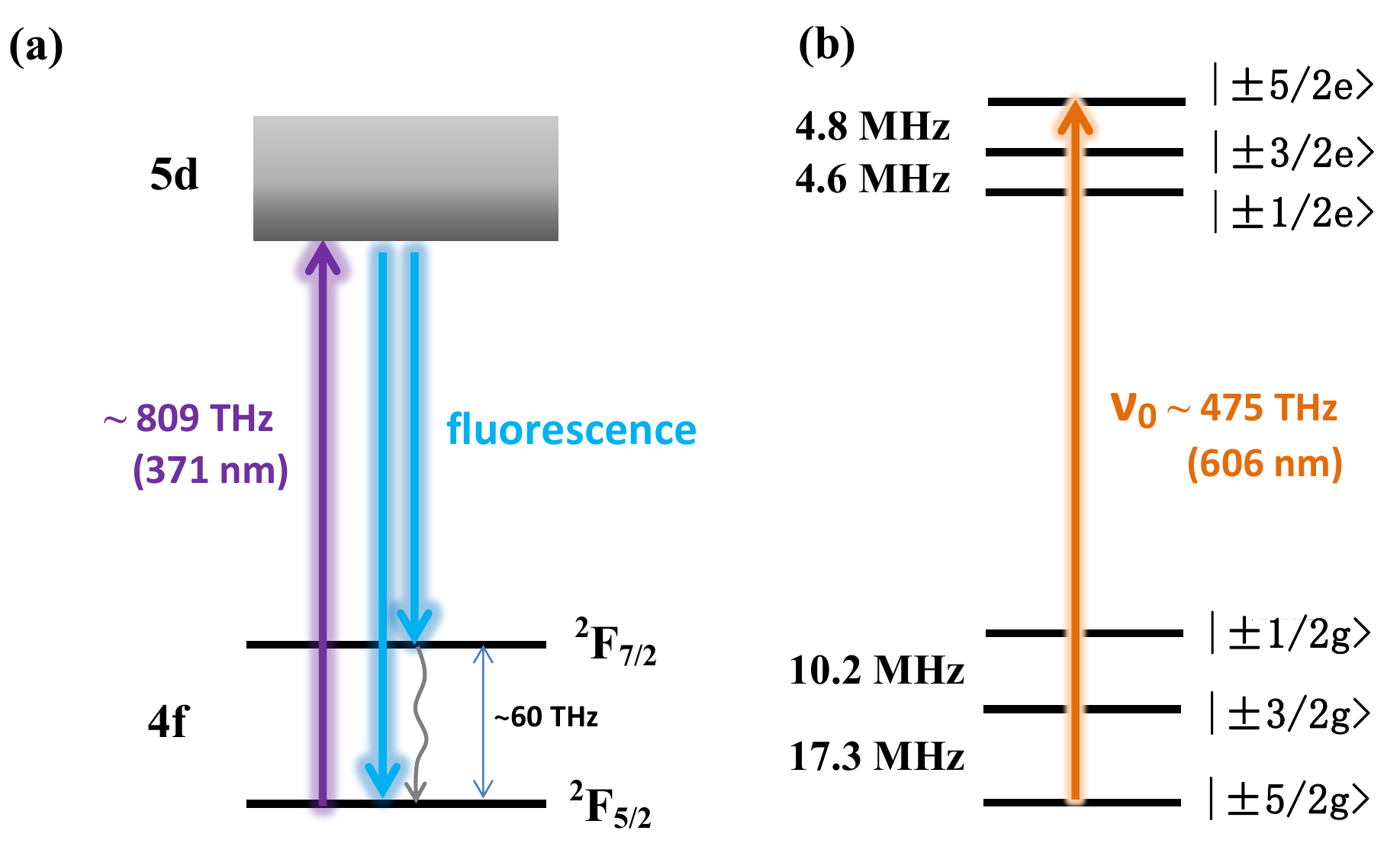}
	\caption{(color online)(a) Schematic level structure for the Ce ion in an Y$_2$SiO$_5\,$ crystal. (b) Schematic level structure for the Pr ion $^{3}H_{4}$ $\rightarrow$ $^{1}D_{2}$ transition in an Y$_2$SiO$_5\,$ crystal. }
	\label{fig.2}
\end{figure}

\begin{center}
\textbf{II. DETERMINATION OF SPECTROSCOPIC PARAMETERS OF Ce IONS} \\
\end{center}

We focused on the parameters that are of special interest to the single ion readout scheme in REIQC.\\

\textbf{A. Zero-phonon line (ZPL) of Ce ions in an Y$_2$SiO$_5$ crystal}\\

The absorption line of interest is the 4f-5d transition of Ce ions as illustrated in \Fig2 (a). Since the 5d levels are less shielded from the environment than the 4f levels, a 4f-5d transition are often largely broadened by the external perturbations, for instance, defects in the crystal or electron-phonon coupling to the crystal lattice. The experiments were carried out at 2 K to greatly reduce the phonon broadening influence. In this and the following experiments, an external cavity diode laser in a Littrow configuration was used as an excitation source.\\
\begin{figure}[h]
		\includegraphics[width=8.3cm,height=6cm]{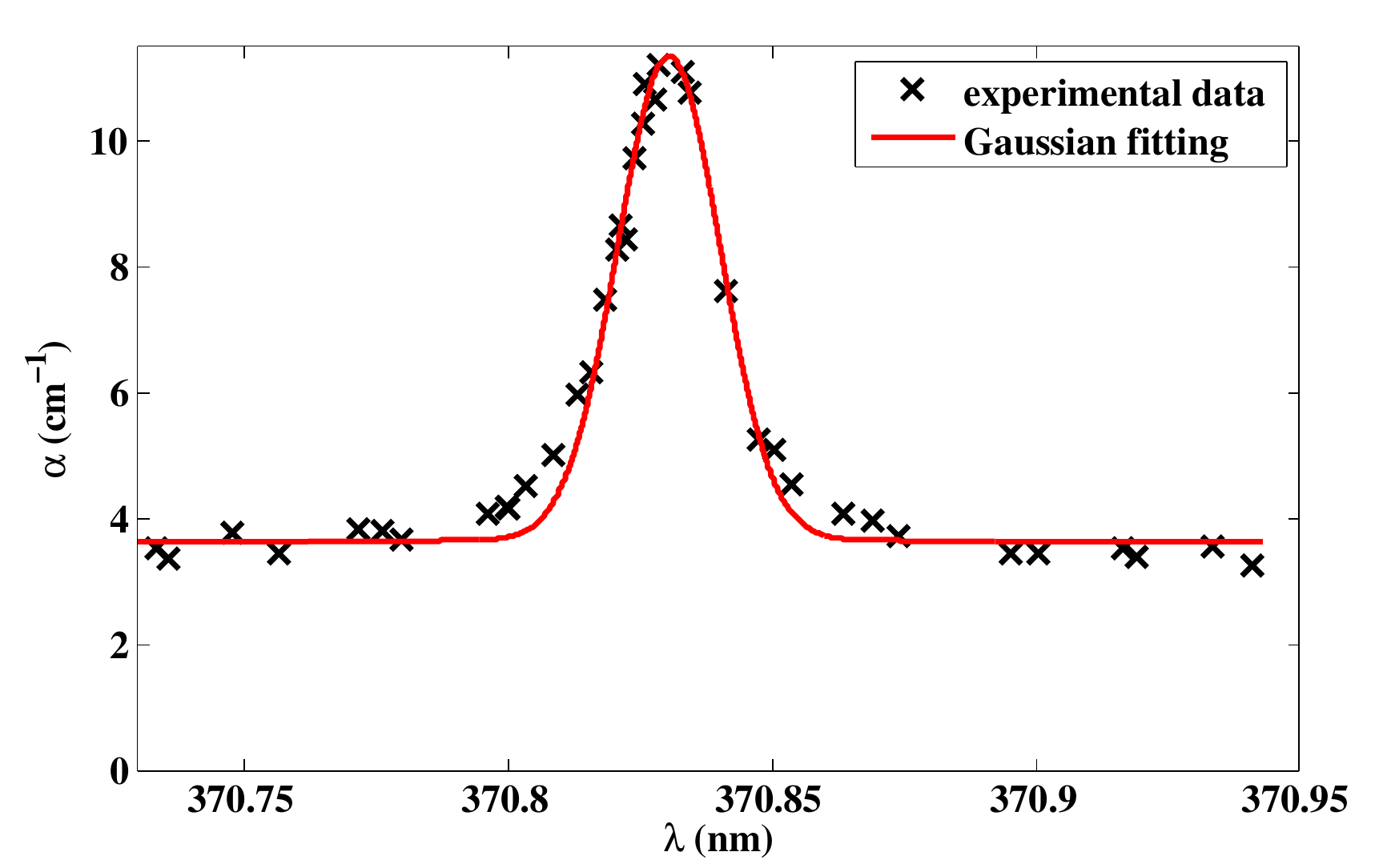}
	\caption{(color online) The inhomogeneous ZPL of \Ce (Site 1) in an Y$_2$SiO$_5\,$ crystal. The line width is about 50 GHz. $\alpha$ is the absorption coefficient. }
	\label{fig.3}
\end{figure}

The inhomogeneous ZPL of \Ce was measured on an Y$_2$SiO$_5$ crystal with a nominal dopant concentration of 0.088 at.$\:$\% relative to the yttrium ions. The result is shown in \Fig3. Crosses are the experimental data and the solid curve is a Gaussian fit. The measured inhomogeneous line width is about 50 GHz (full width at half maximum of the absorption coefficient) with the line center at 370.83 nm. No significant polarization dependence was observed for the absorption. The frequency integrated absorption cross section was obtained from \Fig3 and further the oscillator strength of the transition was calculated to be $\sim$ 9$\times10^{-7}$ by making use of the relation between them as shown in article \cite{Hilborn}. The ZPL shown in \Fig3 sits on a background absorption with $\alpha$ $\simeq$ 3.6 cm$^{-1}$, which most likely comes from the absorption by Ce ions in site 2. More information about this is provided by a fluorescence spectrum with an online excitation (at 370.83 nm) and an offline exciation (at 371.53 nm), shown as the solid and dashed curve in \Fig4. The solid (dashed) curve matches reasonably well with the spectrum from a site 1 (site 2) excitation, shown in \cite{Drozdowski}\cite{Suzuki}. However the solid curve contains an extra shoulder sitting around 435 nm, the origin of which is unknown to us. In the rest of the article, all calculations refer to the ions in site 1, with the site 2 contribution being subtracted as a background. \\

\begin{figure}[h]
		\includegraphics[width=8.2cm,height=5.8cm]{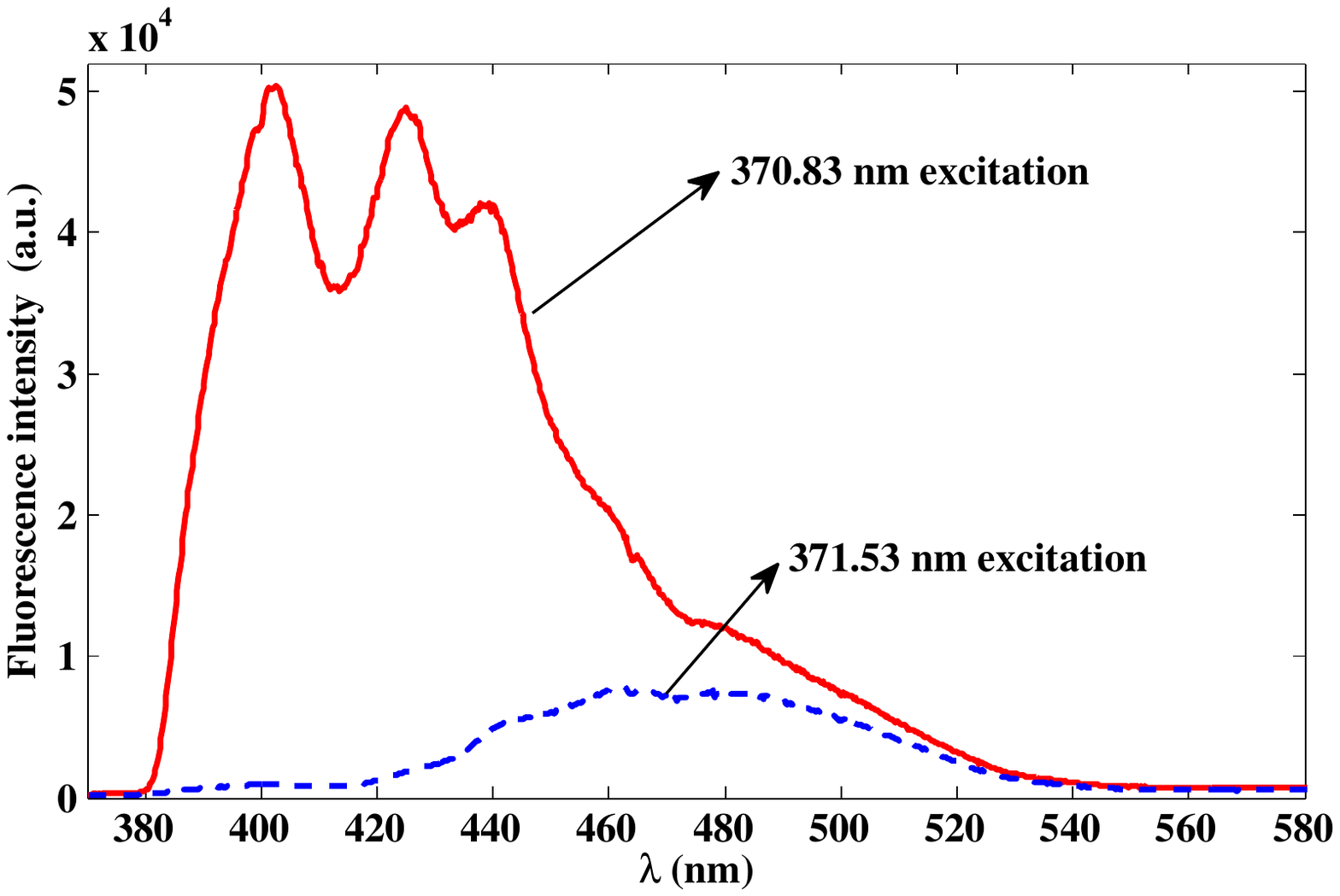}
	\caption{(color online) Fluorescence spectrum of the \Ce (in the same crystal as used for Fig.\:3) from 370.83 nm excitation (online of site 1) (solid line) and 371.53 nm excitation (offline of site 1) (dashed line).}
	\label{fig.4}
\end{figure}

\textbf{B. Homogeneous line width of the ZPL}\\

The homogeneous line width was measured by intensity modulated saturation spectroscopy, where both the pump and probe beams were generated from the external cavity diode laser by a beam splitter (70:30). Both beams were focused by a 200 mm focal length lens (focal diameter $\sim$ \SIunits{85} {\micro\meter}) onto the Ce:Y$_2$SiO$_5$ crystal, which was immersed in liquid helium. The probe beam propagated at an angle of $~3$ degrees relative to the pump beam in order to separate the two beams for the detection. Two Acousto Opical Modulators (AOMs) were used in series in the probe beam line to detune the probe beam frequency within a $\pm$15 MHz range relative to the pump beam frequency. The first one provided a +280 MHz frequency shift, and the second one provided a -265 to -295 MHz frequency shift. The probe beam position movement caused by the AOM at different detuning frequencies was compensated by slightly adjusting the mirror in front of the cryostat to overlap the two beams by maximizing the probe beam transmission. The probe beam was monitored by a photodiode after the cryostat. However, it is hard to directly detect an increase of the transmitted power caused by the saturation of the pump beam, since the top-hat pump beam intensity is only of the level of one percent of the estimated saturation intensity. To improve the signal to noise ratio, a chopper wheel was used to produce an intensity modulation to both the pump and probe beam with modulation frequencies of 302 Hz and 362 Hz, respectively. Due to the nonlinear interaction, the transmitted probe beam not only has the primary frequency modulation but also has an intensity modulation at the sum (664 Hz) and difference frequency (60 Hz). The signal strength at 664 Hz was extracted from the Fourier transform of the transmitted power of the probe beam and recorded as a function of the detuning frequencies between the pump and probe beams, $\bigtriangleup$f. \\

In the experiment, the signal at 664 Hz was 16 times as high as the noise floor when $\bigtriangleup$f is zero but only 1.3\% (at maximum) of the signal strength at 362 Hz. The measurement data is shown in \Fig5, where the vertical axis shows the sum frequency signal normalized by the square of the input probe beam power to compensate for the change in laser power between the data points. \Fig5 reveals a spectral hole width (full width at half maximum) of about 5.8 MHz (with an error bar of $\pm$ 1 MHz caused by the frequency drift of the signal source to AOMs), which is a convolution of the laser line width with the homogeneous line width of the transition. Considering a typical laser linewidth of 0.3 MHz, the homogeneous line width would be about 2.6 MHz ($\pm$ 0.5 MHz). Such a narrow line width can only be achieved when there is no other dephasing channel except the excited state decay, which means the coherence time of the transition is $\sim$ 100 ns.
\begin{figure}[h]
		\includegraphics[width=7.8cm,height=6.0cm]{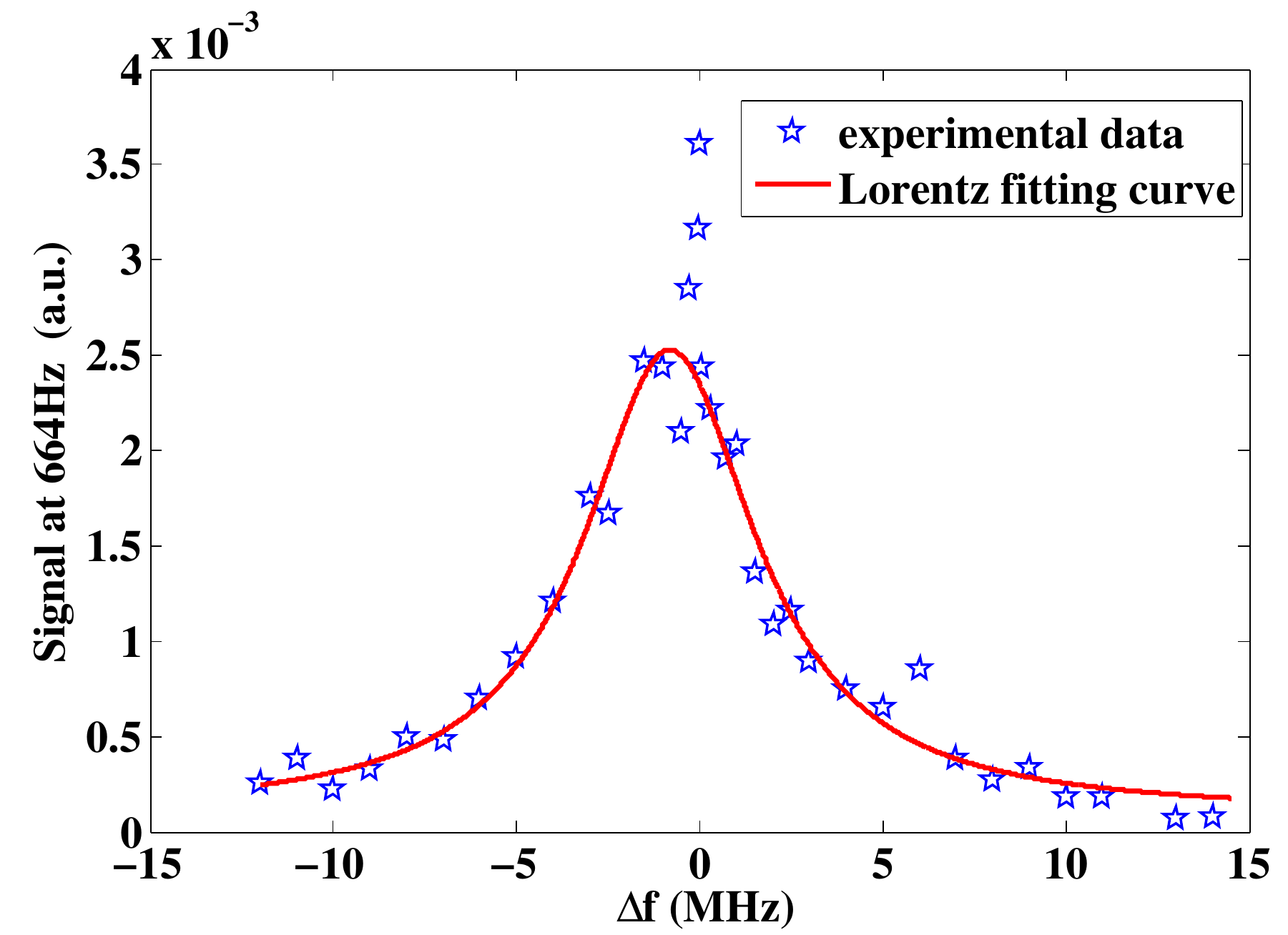}
	\caption{(color online) Ce ion homogeneous line width measured using saturation spectroscopy. The pentagrams are the measured data points. The solid curve is a Lorentz fitting.}
	\label{fig.5}
\end{figure}
Based on the coherence time and the frequency integrated absorption cross section (m$^2\cdot$rad/s) calculated from the inhomogeneous line, the absorption cross section of the Ce ion in site 1 and the corresponding saturation intensity were calculated to be $\sim$ 5$\times10^{-19}\:$m$^{2}$ and $\sim$ 1$\times10^{7}\:$W/m$^{2}$, respectively. The experimental data indicates an increased interaction between the beams within a narrow range ($\sim$ 400 kHz) of frequency detuning close to the line center. The origin of this resonance was not investigated but narrow resonances of this type can occur if the upper state can also decay to a third level which (compared with the relaxation from the excited state) relaxes slowly to the ground state, see \textit{e.g.} \cite{Steel} \cite{Mitsunaga}. The line width of the resonance is then determined by the decay rate of this third level. In the present experiment the third level could be one of the lowest Kramer's doublet on the $^{2}F_{5/2}$ (4f) ground state, comparing with \textit{e.g.} \cite{Kolesov}, and the intrinsic width of the narrow resonance would then be given by the spin-lattice relaxation rate of the ground state. Besides the narrow structure, the existence of a third level also provides a larger saturation effect due to the trapping of ions on this level. The signal at 664 Hz being 0.8\% (not considering the narrow structure) relative to the signal at 362 Hz was observed at zero detuning frequency, but the estimated strength based on the pump beam intensity was $\sim$ 0.1\%. \\

\textbf{C. Ce-Pr interaction}\\

The Ce and Pr ions in Y$_2$SiO$_5$ have different permanent electric dipole moments in their ground and excited states as they sit in non-centrosymmetric sites. This difference in the ground and excited state permanent dipole moment is denoted as $\Delta\mu$. When one ion is changing its state from ground to excited state or vice versa, the surrounding electric field is changed. This permanent dipole-dipole interaction causes a frequency shift, $\Delta f$, of the transition lines of the nearby ions, which depends on the spatial distance, $r$, between those two ions and $\Delta\mu$ of both ions as \cite{Altner}:
\begin{equation}
\Delta f \propto \dfrac{\vec{\Delta\mu_{Ce}}\centerdot\vec{\Delta\mu_{Pr}}}{r^{3}},
\label{dark_def}
\end{equation}
The value of the $\Delta\mu_{Ce}$ of \Ce 4f-5d transition in Y$_2$SiO$_5$ is not previously known. It can \textit{e.g.} be determined by measuring the Stark shift caused by the interaction between the ions and an external electric field \cite{Graf97}. Here we instead implemented a two-pulse photon echo experiment on Pr ions, where Ce ions were excited during the dephasing period of the Pr ions, and the reduction of the echo intensity was observed \cite{Altner}. The reduction is caused by Pr ions, which sit close enough to a Ce ion, to experience a frequency shift when this Ce ion is excited. When this happens the phases of the Pr superposition states evolve at different rates in the dephasing and rephasing periods, which cause the echo intensity to decrease depending on the magnitude of the shift and the evolution time. From the decrease of the echo signal caused by the interaction, $\Delta\mu_{Ce}$ can be calculated \cite{Graf98}.\\

\begin{figure}[h]
		\includegraphics[width=7.9cm,height=2cm]{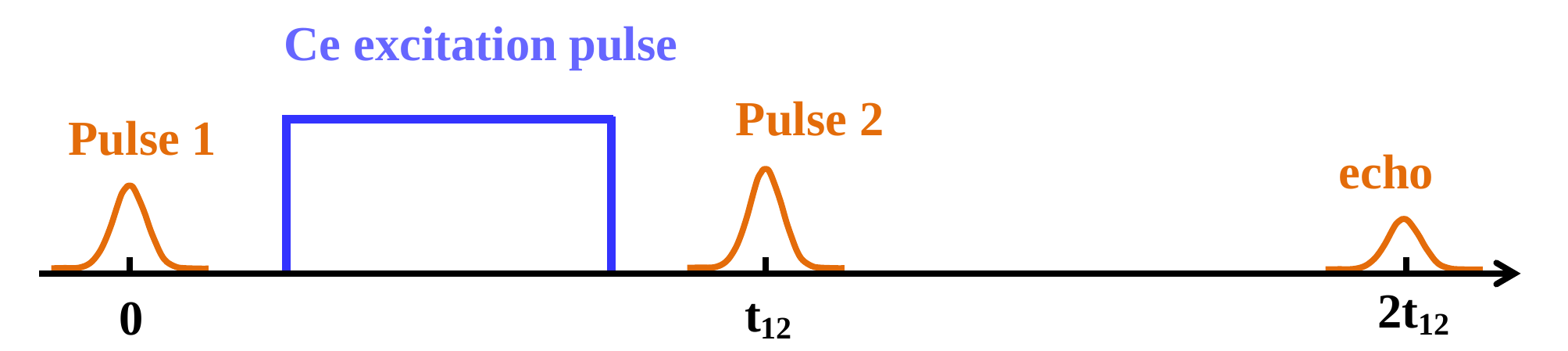}
	\caption{(color online) Pulses sequence for the Ce-Pr interaction measurement.}
	\label{fig.6}
\end{figure}

In the experiment, two weak Gaussian pulses, created by an AOM from a kHz line width 606 nm continuous wave dye laser, with \SIunits{0.15} {\micro\second} duration time ($\sim$ 0.5 mW power) were used to excite the Pr $^3H_{4}\rightarrow^1D_{2}$ transition. The Ce excitation pulse was created from an external cavity diode laser using another AOM. The pulse sequence is shown in \Fig6. Each laser beam was coupled into a single mode fiber for the spatial mode cleaning, and the collimated output beams from the fibers were coupled together by a dichroic mirror with 90\% transmission (reflection) for 606 nm (371 nm). Those two beams were focused onto the crystal with focal diameters about \SIunits{130} {\micro\meter} and \SIunits{170} {\micro\meter} by a low dispersion CaF$_{2}$ lens, respectively. The spatial mode overlap between the two beams within the crystal was ascertained by using a beam profiler to ensure the center of the two beam profiles overlapped with each other within \SIunits{10} {\micro\meter} over the \SIunits{40} {\centi\meter} distance between the dichroic mirror and the common focusing lens. The echo intensity was diffracted by a third AOM into a photomultiplier tube (PMT). Right before the PMT an electronic shutter with \SIunits{100} {\micro\meter} rise time was used to prevent the strong frequency scanning pulses which came after the echo signal going towards the PMT. The frequency scanning pulses were used to remove the persistent spectral holes that would otherwise be created by the two Gaussian pulses over time.\\

The echo intensity from the Pr ions was recorded as a function of the separation time, $t_{12}$, between the two excitation pulses in two situations: Ce ions being excited or not excited (Ce laser blocked) in between those two pulses. The excitation of Ce ions was implemented at wavelengths of either 370.83 nm (referred to as Online exciation hereafter) or 371.54 nm (referred to as Offline excitation). It should be noted that Offline excitation is only offline for the Ce ions in site 1 as discussed in Section II A. The experimental data on the Pr:Ce:Y$_2$SiO$_5$ crystal (grown by Shanghai Institute of Ceramics in China) with 0.05$\%$ of Pr and 0.088$\%$ of Ce dopant concentration relative to the Y ions, is shown in \Fig7. \\
\begin{figure}[h]
		\includegraphics[width=8.2cm,height=5.7cm]{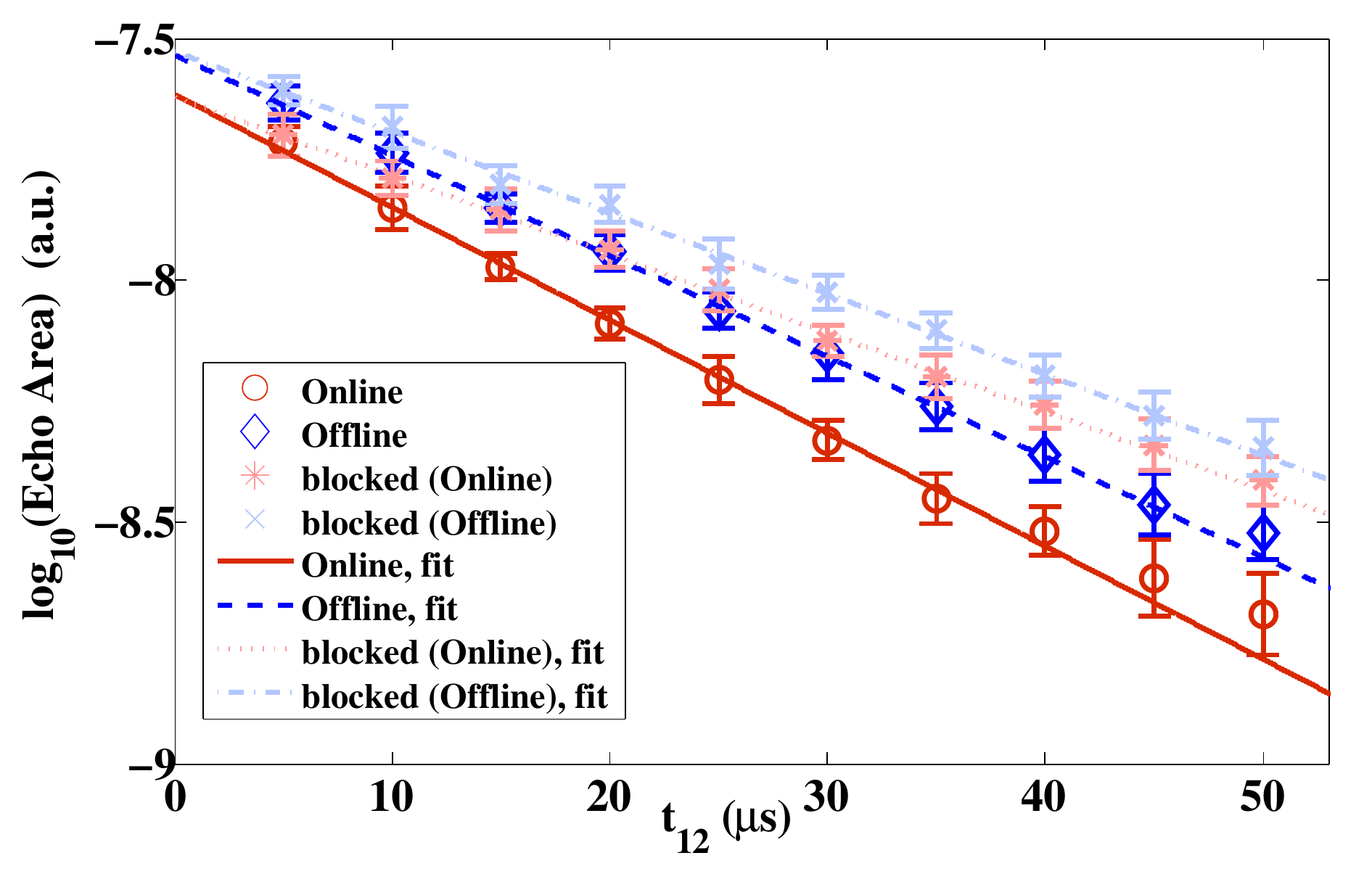}
	\caption{(color online) Decay curve of the echo from the Pr ions on the Pr:Ce:Y$_2$SiO$_5$ crystal. The circles (diamonds) are the experimental echo signal when the Ce ions were excited at an online (offline) wavelength. Error bar shows the standard deviation. The solid, dashed lines are logarithmic fits corresponding to those two cases. The stars (crosses) are the experimental data when the Ce excitation laser was blocked. The difference between them is explained in the text.}
	\label{fig.7}
\end{figure}
\\
Four series of data were recorded. The circles are the echo signals from the Pr ions when Ce ions was excited Online (excitation power $\sim$ 0.5 mW) within time $t_{12}$. Each point is an average value of $\sim$ 30 shots, and the error bar shows the standard deviation. To verify the experimental data, a complimentary data point corresponding to each circle, shown as the stars, was recorded as well with the Ce excitation laser being blocked. At each $t_{12}$, all shots of the circle point were recorded first then followed by the shots of the star point. After recording all data points at various $t_{12}$, the Ce laser was tuned to the Offline position. The corresponding set of data was recorded, shown by the diamonds (Ce laser Offline) and crosses (Ce laser blocked). The solid, dotted, dashed and dash-dotted lines are the logarithmic fit of the experimental data for the cases when the Ce laser is Online, blocked, Offline and blocked, respectively. The two blocked cases are equivalent, but recorded with an offset on the echo signal possibly caused by the laser power variation. However, the slopes are the same within the margin of error, as expected. The echo intensity as a function of the pulse separation time is
\begin{equation}
I_{echo} = I_{0}\centerdot \textit{e}^{-4t_{12}/T_{2}},
\label{dark_def}
\end{equation}
where $I_{0}$ is the maximum echo intensity when extrapolating the separation time to zero. From the data the Pr coherence times of \SIunits{74} {\micro\second} ($\pm$\SIunits{5} {\micro\second} with 70\% confidence interval, the same for the rest) and \SIunits{107} {\micro\second} for the Ce ions being excited Online and not excited, respectively; \SIunits{83} {\micro\second} and \SIunits{104} {\micro\second} for the Ce ions being excited Offline and not excited. The homogeneous line width broadening caused by the Ce Online (Offline) excitation, $\Gamma_{br}$, is 1.2 kHz (0.7 kHz) with $\pm$0.14 kHz for a 70\% confidence interval, which means that the broadening contributing from the ions in site 1 is 0.5 kHz. This value will be used for calculating $\Delta\mu_{Ce}$.\\

  However the reduction in echo intensity could conceivably be caused by other reasons than a frequency shift resulting from the permanent dipole-dipole interaction. For instance, 1) the Pr ions directly absorb the ultraviolet (UV) photons leading to processes shortening the coherence time, which causes the echo intensity to decrease, although this is unlikely to happen, seen from the literature \cite{Kuleshov}. 2) the energy is transferred from Ce ions to Pr ions so that Pr ions are excited to a higher level, which also can cause an echo reduction. To clarify the influence of suggestion 1) above we did exactly the same measurement as before but on an Y$_2$SiO$_5\,$ crystal with the same Pr dopant concentration but no Ce ions. The result is shown in \Fig8,
\begin{figure}[h]
		\includegraphics[width=8.5cm,height=5.5cm]{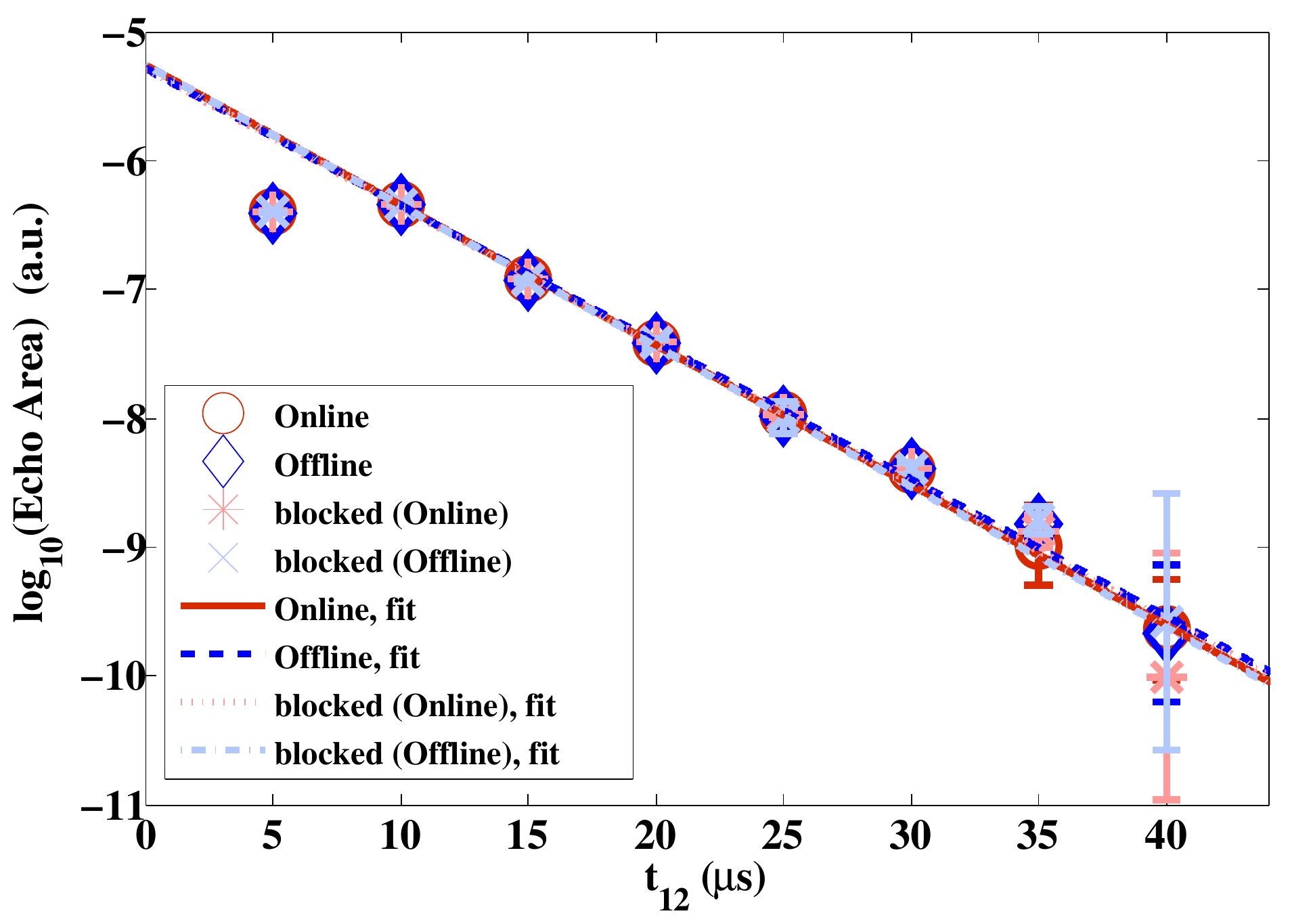}
	\caption{(color online) Echo decay curve on the pure Pr:YSO crystal. Notation of the data symbols are the same as in \Fig7. Circles, diamonds, stars and crosses are for the cases when the Ce laser is Online, Offline, blocked (Online) and blocked (Offline), respectivley. Deviation of the first data point from the fitting line was caused by the saturation of the PMT. }
	\label{fig.8}
\end{figure}
where the Pr homogeneous line widths are the same with and without Ce excitation, which means that the line width broadening shown in \Fig7 does not result from the direct UV photon absorption by the Pr ions. For clarifying whether the interaction can be induced by energy transfer (argument 2) above), we did an excited state population decay measurement on Pr ions in the Pr:Ce:Y$_2$SiO$_5\,$ crystal as following: (i) A zero absorption spectral window centered at frequency, $\nu_{0}$, was created using optical pumping \cite{Nilsson}. (ii) A subset of Pr ions having their transition at frequency $\nu_{0}$ is transferred to the $\vert\pm$5/2g$\rangle$ state (The Pr ion level structure is shown in \Fig2 (b)). (iii) A pulse with pulse area of $\pi$ excites these ions to their $\vert\pm$5/2e$\rangle$ state with an efficiency of more than 85$\%$. (iv) A Ce excitation pulse with duration time, T, was incident on the crystal. If there are Ce-Pr energy transfer or other state changing interactions (not only frequency shifts due to the permanent dipole-dipole interaction) the Pr excited state population should change. (v) After the time, T, the transmission of a pulse scanned in frequency around frequency $\nu_{0}$ (see \Fig2 (b)) determines the population difference between the $\vert\pm$5/2g$\rangle$ and $\vert \pm$5/2e$\rangle$ state by measuring the absorption (this pulse is called the readout pulse in the following passage). Following the procedure above the Pr population difference (normalized to the initial population in $\vert\pm$5/2g$\rangle$ state) between the excited and ground state, $N_{e} - N_{g}$, was recorded as a function of the separation time, T, between the $\pi$ pulse and the readout pulse for the two cases where Ce ions were excited or not excited during the time T.\\

\Fig9 shows that the population differences of Pr ions when Ce ions excited (Online) and not excited (blocked) agree with each other within 3$\%$. Similar measurements from the Ce Offline excitation and Ce laser blocked cases also show the same result. Thus no effect of energy transfer between the Ce and Pr ions was observed. Based on the test on the pure Pr doped crystal (\Fig8) and this excited state population decay measurement of Pr ion (\Fig9), to the best of our knowledge, the homogeneous broadening shown in \Fig7 should be caused by the permanent dipole-dipole interaction between the Pr and Ce ions. \\

\begin{figure}[h]
		\includegraphics[width=8.5cm,height=5.5cm]{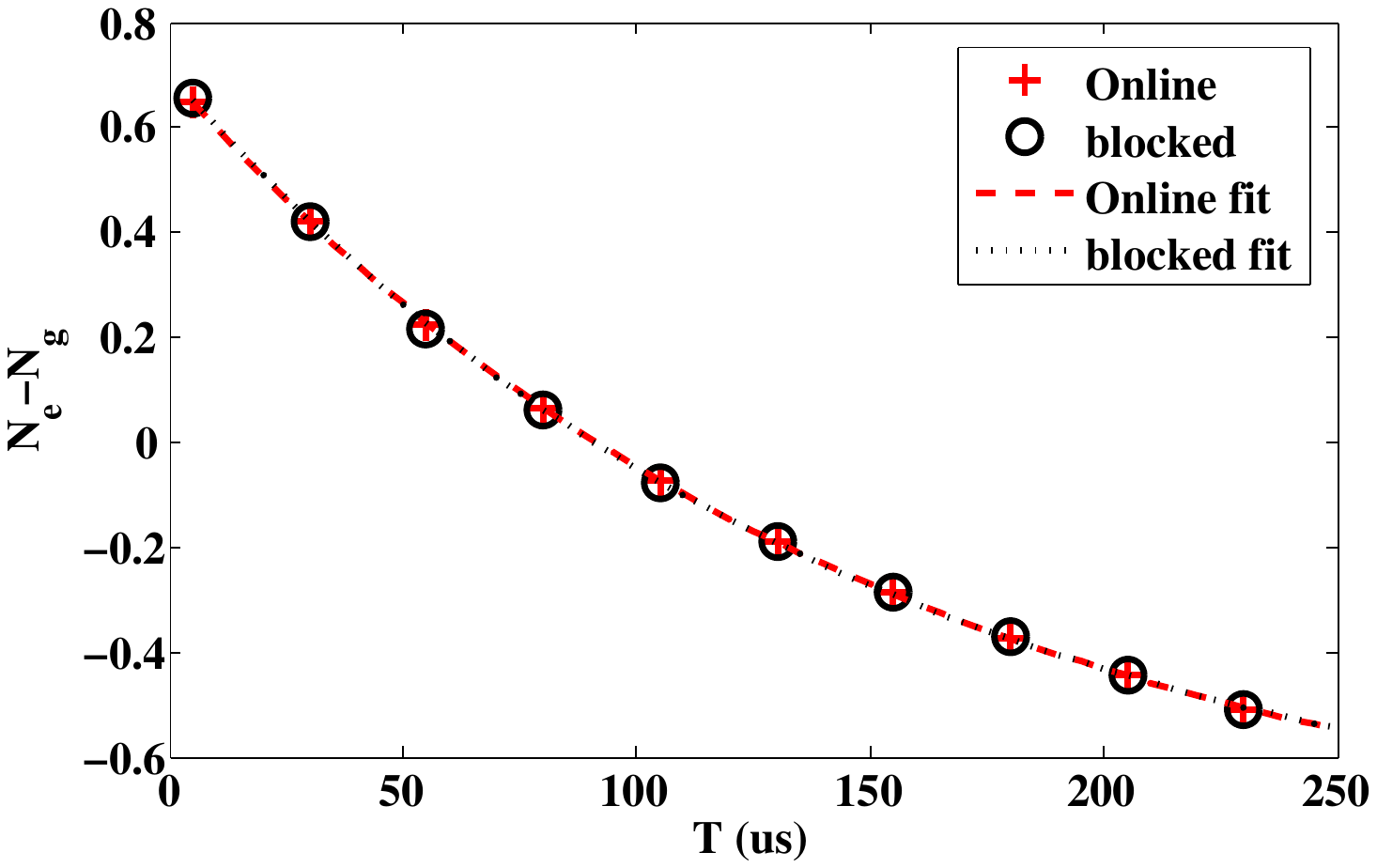}
	\caption{(color online) Population difference between the excited and ground state on the co-doped crystal with Ce Online excitation ($+$) and Ce laser blocked ($\bigcirc$) as a function of time. The dashed and dot-dashed lines are the respective exponential curve fittings. The standard deviation of the data points is $\backsim$ 2\%.}
	\label{fig.9}
\end{figure}

The homogeneous line width broadening induced by the permanent dipole-dipole interaction, $\Gamma_{br}$, relates to the $\Delta\mu$ of the interacting ions in the following \cite{Graf98}:

\begin{equation}
\Gamma_{br} = \dfrac{1}{2\pi} C D_{0}\langle W_{Ce} \rangle
\label{dark_def}
\end{equation}
where
\begin{equation}
C = \dfrac{2}{3}\pi^{2}\textit{p}\langle \vert\kappa\vert\rangle_{\Omega}
\label{dark_def}
\end{equation}
$p$ represents the occupation probability of the Ce ions relative to the total number of ions in the crystal. $\langle \vert\kappa\vert\rangle_{\Omega}$ is of the order of one \cite{Altner}, representing the averaged value of $\kappa$ over all angles, $\Omega$, where $\kappa$ stands for the dipole orientation dependence of the interaction $\kappa = (\hat{\Delta\mu_{Ce}}\cdot\hat{\Delta\mu_{Pr}})-3(\hat{r}\cdot\hat{\Delta\mu_{Ce}})(\hat{r}\cdot\hat{\Delta\mu_{Pr}})$. $\hat{\Delta\mu}$ and $\hat{r}$ are the unit vectors for the permanent dipoles moment change and the displacement, respectively. \\

$D_{0}$ describes the magnitude of the permanent dipole-dipole interaction with a unit distance of $r_{0}$,

\begin{equation}
D_{0} = \eta\left( 0\right)\dfrac{\vec{\Delta\mu_{Ce}}\cdot\vec{\Delta\mu_{Pr}}}{4\pi\epsilon_{0}\hslash r_{0}^{3}}
\label{dark_def}
\end{equation}
where $\vec{\Delta\mu_{Ce}}$  and $\vec{\Delta\mu_{Pr}}$ is the difference of the permanent dipole moment in the ground and excited state of Ce and Pr ions, respectively. $\eta(0)$ = 1.28 representing the dielectric correction factor for long range dipole-dipole interaction \cite{Mahan}. $\epsilon_{0}$ is the vacuum permittivity. $r_{0}$ stands for the length of the cube which one ion occupies on average. $\langle W_{Ce} \rangle$ represents the average excitation probability of the Ce ions contributing to the interaction during the dephasing time. \\

For the crystal used in this experiment, $p\simeq1.92\times10^{-4}$ considering a 87\% occupation on site 1 \cite{GKLiu}. $\langle W_{Ce} \rangle\simeq2.8\times10^{-7}$, which is estimated from the saturation intensity (shown in section II B) and integrated over the inhomogeneous line as in \cite{Graf98}. The excitation intensity used for the estimation is the average value over a volume with the beam radius of \textbf{$\rho_{hwhm}$}, the half width at half maximum of the Pr excitation laser intensity, through the crystal thickness of 1 mm. The projection of $\vec{\Delta\mu_{Ce}}$ (site 1) onto $\vec{\Delta\mu_{Pr}}$ is calculated as $6.3 \times10^{-30}\:$C\cdot m using Eq. (3), (4) and (5) with $r_{0}$ = 0.237 nm, $\langle \vert\kappa\vert\rangle_{\Omega}\simeq0.7$ and $\Delta\mu_{Pr}= 2.43 \times10^{-31}\:$C\cdot m \cite{Graf98}. It is about 26 times as large as $\Delta\mu_{Pr}$. We also calculated $\langle W_{Ce} \rangle$ from the ratio of the number of absorbed photons (equivalently excited Ce ions) during 50 ns over the total number of Ce ions in site 1 for a certain volume (given by the laser beam profile). The value is $2.3\times10^{-7}$, which agrees with the previous value within 20\%. \\

This is an encouraging result indicating that a Pr ion which sits 10 nm (average ion-ion distance with a dopant concentration of 0.05\%) away from a Ce ion in site 1 would shift the Ce transition frequency about 25 MHz once it is excited. This would clearly shift the Ce ion in site 1 out of resonance with the readout laser which is initially on resonance. However, the readout laser is still on resonance with the Ce ions in site 2 within the laser excitation volume since these ions are not excited at their zero-phonon line and the absorption spectrum is broader than the permanent diople-dipole interaction induced shift. This will give us a background fluoresence even when the single Ce ion in site 1 is shifted out of resonance, which will decrease the ON/OFF fluorescence contrast. However based on our preliminary investigation on the single Ce ion (site 1) detection, the fluorescence emitting from the Ce ions in site 1 can still be used as an indicator to show which state ($\vert0\rangle$ or $\vert1\rangle$) the Pr ion occupies. \\

\begin{center}
\textbf{III. CONCLUSION }\\
\end{center}

The spectroscopic properties of \Ce doped in an Y$_2$SiO$_5\,$ crystal were characterized for investigating the possibility to use it as a probe for detecting which hyperfine ground state a nearby ion (\textit{e.g.} Pr) is occupying. Particularly (i) the ZPL of the 4f-5d transition of \Ce doped in Y$_2$SiO$_5\,$ was found around 370.83 nm with a line width of 50 GHz. (ii) The homogeneous line width was measured by intensity modulated saturation spectroscopy to be $\sim$ 3 MHz. It is essentially limited by the excited state lifetime for the \Ce 4f-5d transition, which is the optimal case for using \Ce as a readout ion. In this experiment we also have not observed any signs that there is a long-lived trapping state for the Ce ion, which means that the fluorescence can be cycled as many times as is needed. From the line width measurements the oscillator strength, absorption cross section and saturation intensity were calculated to be $\backsim$ 9$\times10^{-7}$, $\backsim$ 5$\times10^{-19}$ m$^{2}$ and $\backsim$ 1$\times10^{7}$  W/m$^{2}$, respectively. (iii) The difference in the permanent dipole moment for the 4f ($^{2}F_{5/2}$) and lowest 5d states was measured and the Ce-Pr interaction was also demonstrated through a photon echo experiment. The projection of $\vec{\Delta\mu_{Ce}}$ onto $\vec{\Delta\mu_{Pr}}$ was measured to be  $\backsim$ 6.3$\times10^{-30}$ C$\cdot$ m which is about 26 times of $\Delta\mu_{Pr}$ for the Pr ion $^{3}H_{4}\rightarrow^{1}D_{2}$ transition. The data obtained so far shows that the Ce ion is a very promising readout ion candidate and a set-up for single Ce ion detection by observing the 5d-4f fluorescence is currently under construction. Single Ce ion detection in a YAG (\YAG) crystal was recently demonstrated \cite{Kolesov2}. The ability of state selective readout of a single rare-earth ion in inorganic crystals would be a significant step forward for quantum computing in these materials where high fidelity gate operations on ensembles have already been carried out \cite{Longdell}\cite{Rippe}. It also opens the possibility to use rare-earth ions as extraordinarily sensitive probes of the local environment in these types of crystals. The ability to carry out (and read out) operations on individual ions would strongly address the scalability problem and greatly reduce the qubit operation time since fewer and simpler pulses can be used when the dephasing caused by the inhomogeneous broadening is no longer a concern. It is also quite clear that rare-earth ion doped crystals generally have excellent properties for preserving quantum states seen from the impressive quantum memory development that is taking place in these materials \cite{Hedges}\cite{Saglamyurek}\cite{Imam}. \\

\begin{center}
\textbf{Acknowledgement }\\
\end{center}
We thank Prof. Marco Betinelli for the helpful discussion. This work was supported by the Swedish Research Council (VR), the Knut and Alice Wallenberg Foundation (KAW), the Maja och Erik Lindqvists forskningsstiftelse, the Crafoord Foundation and the EC FP7 Contract No. 247743 (QuRep), (Marie Curie Action) REA grant agreement no. 287252 (CIPRIS), Lund Laser Center (LLC) and the Nanometer Structure Consortium at Lund University (nmC@LU).

\bibliographystyle{prsty}

\begin{thebibliography}{99}

\bibitem{Ladd}
T. D. Ladd, F. Jelezko, R. Laflamme, Y. Nakamura, C. Monroe, and J. L. O’Brien,
Nature \textbf{464}, 08812 (2010).

\bibitem{Blatt}
R. Blatt and D. Wineland,
Nature \textbf{453}, 1008 (2008).


\bibitem{Jones}
J. A. Jones,
Prog. NMR Spectrosc. \textbf{59}, 91 (2011).


\bibitem{Plantenberg}
J. H. Plantenberg, P. C. de Groot, C. J. P. M. Harmans1 and J. E. Mooij,
Nature \textbf{447}, 836 (2007).

\bibitem{Nizovtsev}
A. P. Nizovtsev, S. Ya. Kilin, F. Jelezko, T. Gaebal, I. Popa, A. Gruber, and J. Wrachtrup,
Opt. Spectrosc. \textbf{99}, 233 (2005).

\bibitem{Ohlsson}
N. Ohlsson, R. K. Mohan and S. Kr\"{o}ll,
Opt. Commun. \textbf{201}, 71 (2002).

\bibitem{Longdell}
J. J. Longdell and M. J. Sellars,
Phys. Rev. A. \textbf{69}, 032307 (2004).

\bibitem{Rippe}
L. Rippe, B. Julsgaard, A. Walther, Y. Ying, and S. Kr\"{o}ll,
Phys. Rev. A. \textbf{77}, 022307 (2008).

\bibitem{Wesenberg}
J. H. Wesenberg, K. M$\o$lmer, L. Rippe and S. Kr\"{o}ll,
Phys. Rev. A \textbf{75}, 012304 (2007).

\bibitem{Samuel}
S. Bengtsson,
\textit{Simulation and modeling of Rare earth ion based quantum gate operations},
Master thesis, Lund University, Sweden (2012).

\bibitem{Walther}
A. Walther, B. Julsgaard, L. Rippe, Y. Ying, S. Kr\"{o}ll, R Fisher and
S Glaser,
Phys. Scr. \textbf{T137}, 014009 (2009).

\bibitem{Olivier}
Suggested by Olivier Guillot-No$\ddot{e}$l in the European Quantum Information Processing and Computing workshop, Rome, 2004.

\bibitem{Aitasalo}
T. Aitasalo, J. Hölsäa, M. Lastusaari, J. Legendziewicz, J. Niittykoski, and F. Pellé,
Opt. Materials \textbf{26}, 107 (2004).

\bibitem{Julio}
J. E. Hernandez,
Master thesis, Lund University, Sweden (2006).

\bibitem{Hilborn}
R. C. Hilborn,
Am. J. Phys. \textbf{50}, 982 (1982).

\bibitem{Drozdowski}
W. Drozdowski, A. J. Wojtowicz, D. Wi´sniewski,
P. Szupryczy´nski, S. Janus, J. Lefaucheur and Z. Gou,
J. Alloy. Compd. \textbf{380}, 146 (2004).

\bibitem{Suzuki}
H. Suzuki, T. A. Tombrello, C. L. Melcher and J.S. Schweitzer,
Nucl. Instr. and Meth. A \textbf{320}, 263 (1992).

%
\bibitem{Altner}
S. B. Altner, G. Zumofen, U. P. Wild and M. Mitsunaga,
Phys. Rev. B \textbf{54}, 17493 (1996).
%
\bibitem{Steel}
D. G. Steel and S. C. Rand,
Phys. Rev. Lett. \textbf{55}, 2285 (1985).

\bibitem{Mitsunaga}
M. Mitsunaga, N. Uesugi and K. Sugiyama,
Opt. Lett. \textbf{18}, 1256 (1993).

\bibitem{Kolesov}
R. Kolesov,
Phys. Rev. A \textbf{76}, 043831 (2007).

\bibitem{Graf97}
F. R. Graf, A. Renn, and U. P. Wild,
Phys. Rev. B \textbf{55}, 11225 (1997).
%
\bibitem{Kuleshov}
N. V. Kuleshov, V. G. Shcherbitsky, A. A. Lagatskya, V. P. Mikhailov, B.I. Minkov, T. Danger, T. Sandrock, and G. Huber,
J. of Lumin. \textbf{71}, 27 (1997).
%
\bibitem{Nilsson}
M. Nilsson, L. Rippe, S. Kr\"{o}ll, R. Klieber and D. Suter,
Phys. Rev. B \textbf{70}, 214116 (2004).
%
%
%
%
\bibitem{Graf98}
F. R. Graf, A. Renn, G. Zumofen, and U. P. Wild,
Phys. Rev. B \textbf{58}, 5462 (1998).
%
\bibitem{Mahan}
G. D. Mahan,
Phys. Rev. \textbf{153}, 983 (1967).
%
\bibitem{GKLiu}
Y. C. Sun,
in \textit{Spectroscopic Properties of Rare Earths in Optical Materials}, Chapter 7, edited by G. Liu and B. Jacquier, Springer, Berlin (2005).
%
\bibitem{Kolesov2}
R. Kolesov, K. Xia, R. Reuter, R. St\"{o}hr, T. Inal, P. Siyushev, and J. Wrachtrup,
arXiv. org, arXiv: physics / 1301.5215 (2013).
%
\bibitem{Hedges}
M. P. Hedges, J. J. Longdell, Y. M. Li and M. J. Sellars,
Nature \textbf{465}, 1052 (2010).
%
\bibitem{Saglamyurek}
E. Saglamyurek, N. Sinclair, J. Jin, J. A. Slater, D. Oblak, F. Bussieres, M. George, R. Ricken, W. Sohler and W. Tittel,
Nature \textbf{469}, 512 (2011).
%
\bibitem{Imam}
I. Usmani, Ch. Clasusen, F. Bussi\`{e}res, N. Sangouard, M. Afzelius and N. Gisin.
Nature Photonics \textbf{6}, 234 (2012).
%
%
\end{thebibliography}

\end{document}